\documentclass[twocolumn,secnumarabic,amssymb,nobibnotes,aps,prd]{revtex4-1}
\usepackage[spanish]{babel}
\usepackage[T1]{fontenc}
\usepackage{dcolumn}
\usepackage{bm}
\usepackage[utf8]{inputenc}
\usepackage{geometry} 
\usepackage{graphicx} 
\usepackage{sidecap}

\usepackage{color}
\usepackage{float} 
\usepackage{wrapfig} 

\usepackage{lipsum} 
\usepackage{epstopdf}

\graphicspath{{Pictures/}} 
\begin{document}

\title{Design and characterization of a Zeeman Slower}

\author{M. Guevara-Bertsch$^{1,2}$, L. Salfenmoser$^{2,3}$, A. Chavarr\'ia-Sibaja$^{2}$, E. Avenda\~no$^{1,2}$ y O.A. Herrera-Sancho$^{1,2,4}$ }
\address{$^{1}$ Escuela de F\'isica, Universidad de Costa Rica, 2060 San Pedro, San Jos\'e, Costa Rica}
\address{$^{2}$ Centro de Investigaci\'on en Ciencia e Ingenier\'ia de Materiales, Universidad de Costa Rica, 2060 San Pedro, San Jos\'e, Costa Rica}
\address{$^{3}$ Karlsruhe Institute of Technology }
\address{$^{4}$ Institut für Quantenoptik und Quanteninformation, Österreichische Akademie der Wissenschaften, Technikerstr. 21a, 6020}

\date{\today}

\begin{abstract}

\begin{center}
\large{Abstract}
\end{center}
We report on an investigation of a method that applies simultaneously two different mathematical models in order to optimize the design of a Zeeman Slower towards the implementation of ultra cold  atoms in solid state physics. We introduce the implementation of a finite element simulation that allows us to predict with great accuracy the magnetic field intensity profile generated by the proposed design. Through the prediction of the behavior of the Zeeman Slower a greater control is acquired, which allows the optimization of the different experimental variables. We applied the method in the design of a multilayer solenoidal ``Spin-Flip'' Zeeman Slower for strontium atoms. The magnetic intensity profile  generated by the Zeeman Slower is in agreement with the magnetic field strength profile necessary for the atom cooling and tends to zero in both end sides. The latter terms are essential in order to optimize the amount of trapped and cooled atoms. 

\begin{flushleft}


\end{flushleft}

\begin{center}
\large{Resumen}
\end{center}

Nosotros presentamos un m\'etodo que utiliza simult\'aneamente dos modelos matem\'aticos, con el objetivo de optimizar el dise\~no de un desacelerador Zeeman, con miras a la implementaci\'on de \'atomos ultrafr\'ios a la f\'isica del estado s\'olido. Proponemos la implementaci\'on novedosa de una simulaci\'on por medio de elementos finitos con la cual es posible predecir con mucha precisi\'on el perfil de intensidad del campo magn\'etico generado por el dise\~no realizado. Al poder predecir el comportamiento del desacelerador Zeeman se adquiere un mayor control, a partir del cual es posible optimizar las diferentes variables experimentales. El m\'etodo propuesto es aplicado para el dise\~no y construcci\'on de un desacelerador Zeeman solenoidal de tipo ``Spin Flip'' para \'atomos de estroncio. El perfil de intensidades de campo magn\'etico generado por el desacelerador Zeeman construido concuerda con el perfil de intensidades de campo magn\'etico necesario para el enfriamiento de \'atomos de estroncio y tiene adem\'as la ventaja que la intensidad de campo magn\'etico tiende a cero en los extremos. Ambas condiciones permiten incrementar la cantidad de \'atomos enfriados y  atrapados.

\begin{flushleft}

\textsl{Descriptores}: \'atomos ultrafr\'ios, enfriamiento de \'atomos, desacelerador Zeeman

\end{flushleft}

\end{abstract}

\maketitle

\section{Introducci\'on}
\label{intro}

La creaci\'on experimental de cristales a partir de un gas at\'omico representa la culminaci\'on de uno de los sue\~nos m\'as grandes en el estudio de la f\'isica del estado s\'olido. Es as\'i que la posibilidad de acomodar los \'atomos de la forma que se desee para estudiar su comportamiento abre las puertas a una nueva forma de entender los materiales \cite{lattices, solidstate}. La idea anterior fue plausible \'unicamente a partir de la realizaci\'on experimental de sistemas de \'atomos ultrafr\'ios, es decir a partir del momento que se logra llevar los \'atomos a sus niveles m\'as bajos de energ\'ia \cite{ultracold, radiation}. El desarrollo experimental anterior permiti\'o, por ejemplo, la realizaci\'on de condensados de Bose-Einstein, as\'i como la observaci\'on de fen\'omenos c\'uanticos predichos te\'oricamente en las primeras d\'ecadas del siglo XX,  como es descrito en las Refs.~\cite{BEC, ketterle}. Los \'atomos, al tener una energ\'ia cin\'etica reducida, pueden ser atrapados en una red \'optica, con lo cual se adquiere la libertad de ordenarlos en cualquier posici\'on que se desee \cite{Petsas-1994, Greiner-2001, Bloch-2008}. Los desarrollos anteriores permitir\'ian generar, de forma controlada, cualquier arreglo p\'eriodico de \'atomos que se desee estudiar.  

La aplicaci\'on de \'atomos ultrafr\'ios a la rama de la f\'isica del estado s\'olido, conlleva, sin embargo, como es explicado en las Refs.~\cite{Moses:2015, Yan:2013}, un nuevo reto experimental: para poder realizar simulaciones de materiales con \'atomos ultrafr\'ios se requiere que el cristal formado tenga una densidad de \'atomos muy elevada. Esta densidad tiene que ser tal que los \'atomos capturados en la red \'optica interact\'uen en un sistema entrelazado cu\'antico en vez de una colecci\'on de part\'iculas aisladas. Para poder alcanzar la formaci\'on de un sistema entrelazado se busca alcanzar el caso ideal en el cual la cantidad de \'atomos atrapados sea igual o mayor que la cantidad de sitios disponibles en la red \'optica, como es explicado en la Ref.~\cite{Hemmerich:1995}. Hasta la fecha, la mayor ocupaci\'on de \'atomos en una red tridimencional reportada es de aproximadamente 0.3 \cite{Moses:2015}. Para poder alcanzar la alta densidad del cristal mencionada, el primer paso consiste en lograr atrapar y enfriar la mayor cantidad de \'atomos posible. Por lo tanto, es esencial, desde el punto de vista experimental, optimizar cada una de las etapas del sistema de enfriamiento de \'atomos.

 El proceso necesario para poder reducir la energ\'ia cin\'etica de los \'atomos a sus niveles m\'as bajos, de forma general, se realiza en cuatro etapas: (i) desaceleraci\'on de un haz at\'omico, (ii) enfriamiento sub-Doppler y captura en una trampa magneto-\'optica, (iii) transferencia a un potencial conservativo y (iv) enfriamiento evaporativo \cite{ultracold, solidstate}. En la etapa (i) se busca reducir la rapidez de los \'atomos a un orden de magnitud de alrededor de $10 \frac{\textrm{m}}{\textrm{s}}$ para que puedan ser atrapados (etapa (ii))  y transferidos (etapa (iii)) a la trampa magneto-\'optica y finalmente enfriados por evaporaci\'on (etapa (iv)). Por lo tanto, el primer paso para incrementar la cantidad de \'atomos capturados en la trampa magneto-\'optica (etapa (ii)) consiste en reducir la rapidez inicial de la mayor cantidad de \'atomos. Usualmente, para resolver el problema anterior se utiliza un sistema de enfriamiento de \'atomos por medio de un campo magn\'etico variable llamado ``desacelerador Zeeman'' (Zeeman Slower, en ingl\'es) \cite{lasercooling}.

Para desacelerar la mayor cantidad de \'atomos es necesario que el campo magn\'etico generado se adec\'ue lo m\'as posible a las variaciones de la rapidez de los \'atomos en el interior del desacelerador Zeeman  y que adem\'as, los \'atomos no sufran ninguna interferencia con el campo magn\'etico una vez que salen de este. Para resolver este problema existen diversos tipos de desaceleradores Zeeman, cuyas caracter\'isticas var\'ian de acuerdo a las necesidades requeridas por el experimento \cite{solidstate}. Las investigaciones reportan modificaciones en los dise\~nos, entre las cuales se encuentran: variaci\'on de la longitud, diferentes tipos de materiales utilizados, variaciones en el embobinado y la utilizaci\'on de campos sim\'etricos de tipo ``Spin-Flip'' \cite{spinflip, Courtillot-2003, Cheiney:2011}. No obstante, no fue posible encontrar en la literatura alg\'un m\'etodo o t\'ecnica sistem\'atica que permita elegir las propiedades id\'oneas del dise\~no. Por lo tanto, nosotros proponemos un m\'etodo que permite determinar, de forma sencilla y precisa, la configuraci\'on \'optima de cualquier dise\~no de desacelerador Zeeman.

Recientemente, las investigaciones basadas en el enfriamiento de átomos alcalinot\'erreos han atra\'ido particularmente la atenci\'on \cite{Boyd, Jila}. Como es presentado en Ref.~\cite{stellmer-2014, Martinez-2009} estos elementos, al tener dos electrones en su capa externa, tienen la ventaja que sus is\'otopos bos\'onicos cuentan con un estado base cuyo momento magn\'etico es cero y adem\'as presentan tripletes metaestables, lo cual es una ventaja debido a que facilita el uso de transiciones con anchos de l\'inea peque\~no \cite{stellmer-2009}. Al tener estas cualidades, estos elementos son buenos candidatos para la realizaci\'on de experimentos de sistemas de muchos cuerpos altamente correlacionados. Actualmente, debido a estas propiedades, el estroncio se est\'a convirtiendo en uno de los elementos m\'as estudiados para la realizaci\'on de condensados de Bose Einstein, en la realizaci\'on de relojes \'opticos, el estudio de las variaciones de constantes fundamentales as\'i como en la realizaci\'on de simulaciones y computadores c\'uanticos Refs.~\cite{Daley-2008, Gorshkov-2010}.

En esta investigaci\'on reportamos la implementaci\'on de un m\'etodo compuesto por dos modelos matem\'aticos para optimizar y facilitar el dise\~no de un desacelerador Zeeman. Este m\'etodo fue utilizado para realizar el dise\~no y construcci\'on de un desacelerador Zeeman para \'atomos de estroncio, espec\'ificamente para el isotopo naturalmente m\'as abundante ($^{88}{\textrm{Sr}}$) . En la secci\'on II se describen los modelos desarrollados para optimizar la realizaci\'on de un dise\~no de un desacelerador Zeeman. La secci\'on III  describe la construcci\'on del dise\~no propuesto y finalmente en la secci\'on IV se describe la caracterizaci\'on de la intensidad del campo magn\'etico generado por el desacelerador Zeeman construido.

\section{Optimizaci\'on del campo magn\'etico a lo largo del desacelerador Zeeman}
\label{modelos}

El principio de funcionamiento de un desacelerador Zeeman consiste, como su nombre lo indica, en la implementaci\'on del efecto Zeeman para reducir la rapidez de los \'atomos. Al salir de la fuente, los \'atomos tienen una energ\'ia cin\'etica muy elevada, debido a esta raz\'on el primer paso del sistema de enfriamiento consiste en reducir su rapidez en al menos un orden de magnitud. La desaceleraci\'on de un haz de \'atomos se realiza como es explicado en Ref.~\cite{ultracold}, resumidamente, de la siguiente forma: se dirige un haz colimado de \'atomos en direcci\'on opuesta al haz de un l\'aser cuya frecuencia est\'a en resonacia con la transici\'on at\'omica de los \'atomos a enfriar. Los fotones provenientes del l\'aser son absorbidos y r\'apidamente emitidos aleatoriamente por los \'atomos incidentes. Este efecto, en promedio, genera una disminuci\'on de la rapidez de los \'atomos. Sin embargo debido al efecto Doppler conforme los \'atomos se van enfriando, es decir van reduciendo su rapidez, dejan de ``ver'' la frecuencia del l\'aser y consecuentemente dejan de emitir fotones y por lo tanto no son enfriados. As\'i, sin ayuda externa, los \'atomos seguir\'ian a trav\'es del l\'aser con rapidez constante. Lo anterior se puede resolver de dos formas distintas. La primera consiste en variar continuamente la frecuencia del l\'aser por medio del m\'etodo de enfriamiento ``chirping'' \cite{foot}. La segunda t\'ecnica, consiste en la aplicaci\'on de un campo magn\'etico variable que genere una perturbaci\'on  de los niveles de energ\'ia de los \'atomos, debido al efecto Zeeman, cambiando la energ\'ia de la transici\'on at\'omica de forma tal que la frecuencia del l\'aser se encuentre en constante resonancia a lo largo del recorrido de los \'atomos. La t\'ecnica anterior es conocida como ``Desacelerador Zeeman''.

La eficiencia del desacelerador Zeeman depende principalmente de que las variaciones de la intensidad del campo magn\'etico  coincidan con el perfil de intensidades necesario para contrarrestar el efecto Doppler. El perfil de intensidades de campo magn\'etico se regula de tal forma que las transiciones \'atomicas de los \'atomos est\'en siempre en resonancia con la frecuencia del haz del l\'aser \cite{Phillips-1982}. El primer paso en la realizaci\'on de un desacelerador Zeeman consiste en determinar las variaciones de la rapidez de los \'atomos a lo largo del desacelerador Zeeman y a partir de estas variaciones obtener el perfil de campo magn\'etico necesario para mantener a los \'atomos en constante resonancia con la frecuencia del l\'aser.

La fuerza de dispersi\'on necesaria para desacelerar los \'atomos es determinada por la tasa de fotones absorbidos que imparten momentum al \'atomo, $F_{scatt}=\frac{\hbar k \Gamma}{2}$, donde $\hbar$ es la constante de Planck \textit{h}, dividida entre $2\pi$, \textit{k} el n\'umero de onda y $\Gamma$ el ancho de banda natural de la transici\'on utilizada, como es explicado en Ref.~\cite{tesis}. Esta fuerza determina la magnitud de la aceleraci\'on, $\alpha$, con la cual son frenados los \'atomos:
\begin{equation}
\alpha=\frac{ \hbar  \Gamma k}{2m},
\label{eq:3}
\end{equation}
donde \textit{m} es la masa de los \'atomos que se desean enfriar. Por lo tanto la rapidez de los \'atomos en cada punto del dispositivo  experimental \textit{v(z)} se puede obtener simplemente a partir de la siguiente ecuaci\'on:

\begin{equation}
v(z)=\sqrt{v^{2}_{0}-2\alpha z},
\label{eq:2}
\end{equation}
donde $v_{0}$ es la rapidez inicial con la cual ingresan los \'atomos en el enfriador y \textit{z} la posici\'on de los \'atomos  lo largo del enfriador.
Para asegurar absorci\'on continua de los fotones por los \'atomos se aplica un campo magn\'etico \textit{B(z)}, el cual genera una perturbaci\'on a los estados energ\'eticos del \'atomo. De esta forma, se busca que el campo magn\'etico var\'ie de acuerdo a los cambios en la rapidez de los \'atomos \textit{v(z)} y as\'i contrarrestar el efecto Doppler, es decir que var\'ie de acuerdo a la siguiente ecuaci\'on:

\begin{equation}
\omega_{0}+ \frac{\mu_{B}B(z)}{\hbar}=\omega + kv(z),
\label{eq:4}
 \end{equation}
donde $\omega_{0}$ corresponde a la frecuencia de la transici\'on del \'atomo en reposo, $\omega$ la frecuencia del l\'aser, $\mu$ es el magnetr\'on de Bohr y \textit{v(z)} corresponde a la rapidez de los \'atomos a lo largo de la direcci\'on \textit{z} (escogida como la direcci\'on de propagaci\'on a trav\'es del desacelerador Zeeman). Se reemplaza la Ecuaci\'on \ref{eq:2} en la \ref{eq:4} y se obtienen las variaciones del campo magn\'etico a lo largo del enfriador en la direcci\'on\textit{ z}:
\begin{equation}
B(z)=\frac{\hbar }{\mu}\left(\Delta_{0}+ k\sqrt{v^{2}_{0}-2\alpha z}\right) ,
\label{eq:1}
\end{equation}
donde $\Delta_{0}$ corresponde a la diferencia entre la frecuencia del l\'aser y la frecuencia en resonancia de la transici\'on del \'atomo en reposo. Las propiedades espec\'ificas del dise\~no experimental del enfriador de \'atomos dependen del \'atomo que se desee enfriar, pues como se observa en la Ecuaci\'on \ref{eq:1} es funci\'on de las propiedades at\'omicas de los niveles energ\'eticos. A partir de la Ecuaci\'on~\ref{eq:1} es posible entonces determinar el perfil de intensidades del campo magn\'etico necesario en funci\'on del experimento que se desee realizar. 

Para determinar el dise\~no \'optimo del desacelerador Zeeman el m\'etodo propuesto inicia con la caracterizaci\'on  del perfil de intensidades del campo magn\'etico necesario. Para ello se establecen las propiedades especificas del \'atomo que se desee enfriar: masa ``\textit{m}'', ancho de banda natural de la transici\'on``$\Gamma$'', la rapidez inicial ``\textit{$v_{0}$}'' a la que van a ingresar los \'atomos al enfriador y la rapidez final ``\textit{$v(z=l)$}'', a la que se desea llevar a los \'atomos al final del enfriador de longitud ``\textit{l}''. En esta secci\'on del modelo se puede incorporar, de ser necesario, un corrimiento $\Delta_{0}$ en la frecuencia. A partir de estas especificaciones se modela en Mathematica$^{\textregistered}$ de forma sencilla la Ecuaci\'on \ref{eq:1} y se obtiene un gr\'afico de la variaci\'on del campo magn\'etico en funci\'on de la posici\'on a lo largo del enfriador.

Se aplica la primera fase del m\'etodo propuesto al caso del enfriamiento de \'atomos de estroncio. Para este elemento la transici\'on entre el estado base $ ^{1}S_{0}$ y el primer estado excitado $ ^{1}P_{1}$ puede ser utilizada para reducir la rapidez del haz de \'atomos en el interior del desacelerador Zeeman pues es una transición de tipo dipolo que tiene una probabilidad de transición muy alta. La notaci\'on espectrosc\'opica para los niveles energ\'eticos es la siguiente: $^{2S+1}L_{J}$, donde \textsl{L} es el momento angular orbital c\'uantico, \textit{S} es el momentum angular de spin total y \textit{J} el momento angular total c\'uantico. Dicha transici\'on tiene una longitud de onda $\lambda= 461$\,nm y la transici\'on tiene un ancho de banda natural de $\Gamma=2\pi\times32$\,MHz \cite{zeemanslower}. La longitud del enfriador y el perfil de intensidad del campo magn\'etico dependen de la rapidez m\'axima de los \'atomos al salir del horno y de la rapidez final a la que se busca llevarlos. En nuestro caso la fuente de \'atomos utilizada calienta los \'atomos a una temperatura de $500\,^{\circ}\textrm{C}$ esto hace que salgan con una rapidez de aproximadamente $420\,\frac{\textrm{m}}{\textrm{s}}$, buscamos reducir la rapidez a una magnitud de aproximadamente $30\,\frac{\textrm{m}}{\textrm{s}}$.

Al aplicar el modelo utilizando los par\'ametros especificados anteriormente se obtiene que, para reducir la rapidez de los \'atomos de estroncio es necesario generar una intensidad de campo magn\'etico que var\'ia con un rango m\'aximo de 600\,G, el cual concuerda con los valores reportados en Ref.~\cite{zeemanslower}. Para facilitar el embobinado y reducir los efectos de calor de Joule se eligi\'o utilizar un arreglo de campo bipolar, es decir se separ\'o el perfil de intensidad del campo magn\'etico en dos secciones que generan campos magn\'eticos opuestos de igual magnitud, es decir $\left|\textrm B_{min}\right| \approx \left|\textrm B_{max} \right|$, para esto es necesario aplicar un corrimiento en la frecuencia $\Delta_{0}=-2\pi\times 500\,\textrm{MHz}$. Este tipo de arreglo tiene adem\'as la ventaja de que permite reducir el campo magn\'etico en los extremos del desacelerador Zeeman, as\'i mismo permite disminuir los efectos del campo magn\'etico en los \'atomos una vez que salen del desacelerador Zeeman. Al reducir los efectos del campo magn\'etico en la salida del desacelerador Zeeman es posible acercar la trampa magneto-\'optica al desacelerador, mejorando as\'i la eficiencia de captura. Las propiedades que se tienen que cumplir para la realizaci\'on del dise\~no se resumen en el Cuadro \ref{cuadro:2}.

\begin{figure}
\centering
\includegraphics[scale=0.15]{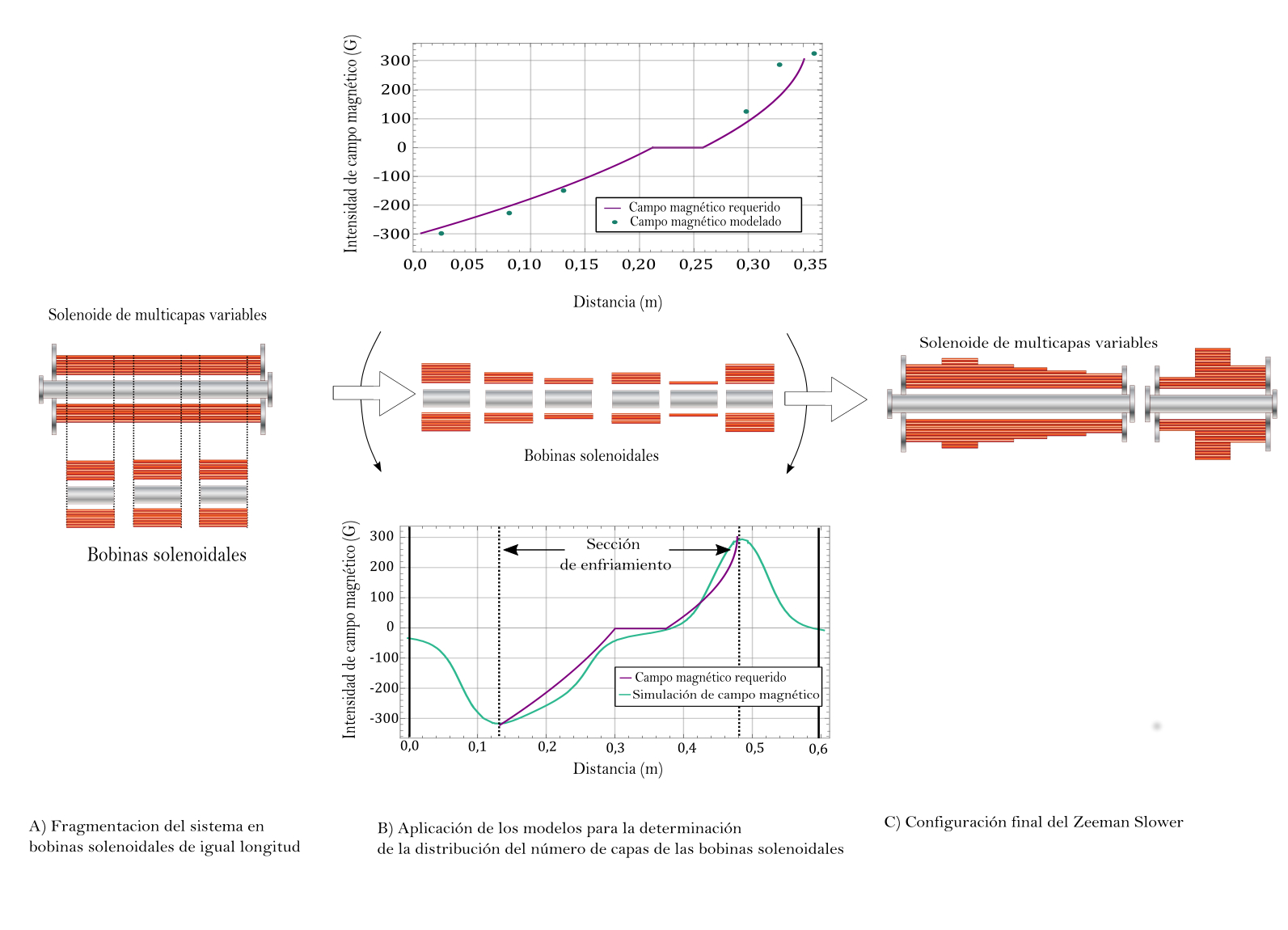}
\caption{Diagrama del m\'etodo utilizado para optimizar la eficiencia de captura de \'atomos del desacelerador Zeeman. A) Se divide el solenoide en una serie de bobinas solenoidales de igual longitud. B) Se busca encontrar la distribuci\'on del n\'umero de capas por bobina aplicando los dos modelos simult\'aneamente. C) Se obtiene la configuraci\'on final del n\'umero de capas en el desacelerador Zeeman.}
\label{modelo}
 \end{figure}

\begin{table}
\begin{tabular}{|l | c |}
\hline
Longitud & 0.6 m \\
Ancho de banda  & 201 MHz\\
Detuning del laser & -3141 MHz \\
Rapidez de captura & 420 m/s\\
Rapidez final & 32 m/s\\
\hline
\end{tabular}
\caption{Especificaciones para el dise\~no del desacelerador Zeeman}
\label{cuadro:2}
\end{table}
A partir de las propiedades definidas en el Cuadro~\ref{cuadro:2} se obtiene el perfil de campo magn\'etico requerido para reducir la rapidez de los \'atomos de estroncio. La gr\'afica obtenida se puede observar en la parte B) de la Figura~\ref{modelo}. Una vez obtenido el perfil de campo magn\'etico requerido se procede con la segunda parte del m\'etodo, la cual consiste en determinar las propiedades del enfriador necesarias para generar dicho campo.

La segunda parte del m\'etodo consiste en en determinar la configuraci\'on del dise\~no para generar el campo magn\'etico requerido. El m\'etodo se aplica para el caso de un dise\~no de tipo solenoide con multicapas variables, similar al empleado en Refs.~\cite{Kondo-1997, Prodan-1982,Phillips-1982}.  La utilizaci\'on de un solenoide tiene varias ventajas con respecto a la utilizaci\'on de un im\'an permanente. Como es explicado en Ref.~\cite{spinflip} una de las ventajas consiste en que este tipo de dise\~no permite encender y apagar la intensidad del campo magn\'etico en cualquier momento. La utilizaci\'on de un solenoide ofrece adem\'as, la posibilidad de modificar el perfil de intensidades generado variando \'unicamente la corriente utilizada. El perfil de intensidad del campo magn\'etico generado por el solenoide depende de la corriente empleada y  de la distribuci\'on de las capas a lo largo del desacelerador Zeeman pues una mayor cantidad de capas en un punto dado genera una intensidad de campo magn\'etico mayor. 

Para determinar la distribuci\'on \'optima del n\'umero de capas en funci\'on de la distancia se utilizaron dos modelos. El primer modelo c\'alcula te\'oricamente la distribuci\'on del n\'umero de capas para obtener el perfil de intensidades buscado. El segundo modelo realiza una simulaci\'on por medio de elementos finitos de la intensidad del campo magn\'etico generada a partir de la distribuci\'on de capas especificada en el primer modelo. Como se observa en la Figura\,\ref{modelo}, en A) se comienza fraccionando la bobina solenoidal de multicapas en una serie de bobinas solenoidales consecutivas de igual longitud cuyo n\'umero de capas es variable. Antes de iniciar el modelo es necesario especificar: el n\'umero de bobinas, la longitud y el diam\'etro del solenoide, as\'i como el diam\'etro y el material del cable que ser\'a utilizado para realizar el embobinado. Estos par\'ametros pueden ser variado y as\'i obtener distintas configuraciones finales. Luego en B) se procede a determinar la distribuci\'on del n\'umero de capas a partir de los dos modelos. Finalmente se obtiene en C) la configuraci\'on final del dise\~no experimental del desacelerador Zeeman. Ambos modelos pueden ser obtenidos libremente en la siguiente p\'agina: $https://github.com/Gustavroot/zeeman\underline{\ }slower$ y se encuentran disponibles para cualquier mejora. 

El primer modelo, realizado en el software Mathematica$^{\textregistered}$, permite obtener la distribuci\'on del n\'umero de capas en cada bobina. El modelo calcula te\'oricamente la contribuci\'on al campo magn\'etico de cada bobina en funci\'on del n\'umero de capas y compara el perfil de intensidad de campo magn\'etico obtenido con el perfil de intensidad de campo magn\'etico requerido obtenido anteriormente. A partir de esto se obtiene el n\'umero de capas necesario en cada bobina. Esto se observa en el gr\'afico superior de la Figura~\ref{modelo}, en B): cada punto verde corresponde al campo generado por una bobina del solenoide con un n\'umero de capas espec\'ifico. Este modelo es similar al reportado en Ref.~\cite{tesis}. El aporte novedoso propuesto con el cual se puede aumentar la eficiencia del desacelerador as\'i como la captura en la trampa magn\'etica corresponde a la implementaci\'on del segundo modelo.

El segundo modelo, desarrollado en el software Comsol$^{\textregistered}$, simula por elementos finitos el perfil de intensidad de campo magn\'etico generado a partir de la distribuci\'on de capas por bobina determinada anteriormente. En este modelo es necesario especificar las variables f\'isicas experimentales propias del dise\~no: distribuci\'on del n\'umero de vueltas por bobina, materiales utilizados, corriente y las dimensiones de cada secci\'on. El modelo simula el comportamiento de un solenoide de capas variables. A partir de los par\'ametros definidos se obtiene una simulaci\'on muy cercana a la realidad del perfil de intensidades del campo magn\'etico en funci\'on de la distancia a lo largo del solenoide, es decir \textit{B(z)} en la Ecuaci\'on~\ref{eq:4}. El gr\'afico inferior de la Figura~\ref{modelo} en B) ilustra el perfil simulado utilizando el segundo modelo.

Para obtener la distribuci\'on del n\'umero de capas en cada bobina se iteran ambos modelos variando los diferentes par\'ametros. Se busca obtener la mejor concordancia posible entre el modelo te\'orico y la simulaci\'on, esto hasta lograr obtener el perfil de intensidad de campo magn\'etico simulado m\'as cercano al perfil de intensidad de campo magn\'etico requerido. A partir de estos dos modelos se obtiene finalmente en C) la configuraci\'on final de la distribuci\'on de capas a lo largo del desacelerador Zeeman. Una de las mayores ventajas observadas al aplicar este met\'odo es que permite controlar de forma muy sencilla la variaci\'on del campo en los extremos del enfriador. Las simulaciones por medio de elementos finitos ofrecen adem\'as la incre\'ible ventaja de que es posible conocer de una forma m\'as cercana a la realidad el comportamiento de un dise\~no experimental.

El m\'etodo se aplic\'o al dise\~no del desacelerador para \'atomos de estroncio. Como observamos en los gr\'aficos de la Secci\'on B) de la Figura~\ref{modelo}, la aplicaci\'on de ambos modelos permiti\'o obtener una distribuci\'on del embobinado que resuelve simult\'aneamente dos necesidades: la generaci\'on de un perfil de intensidad de campo magn\'etico que se ajuste al perfil necesario para contrarrestar el efecto Doppler y que paralelamente se cancele en los extremos del desacelerador.

\section{Montaje Experimental}
\begin{figure}
\centering
\includegraphics[height=4cm]{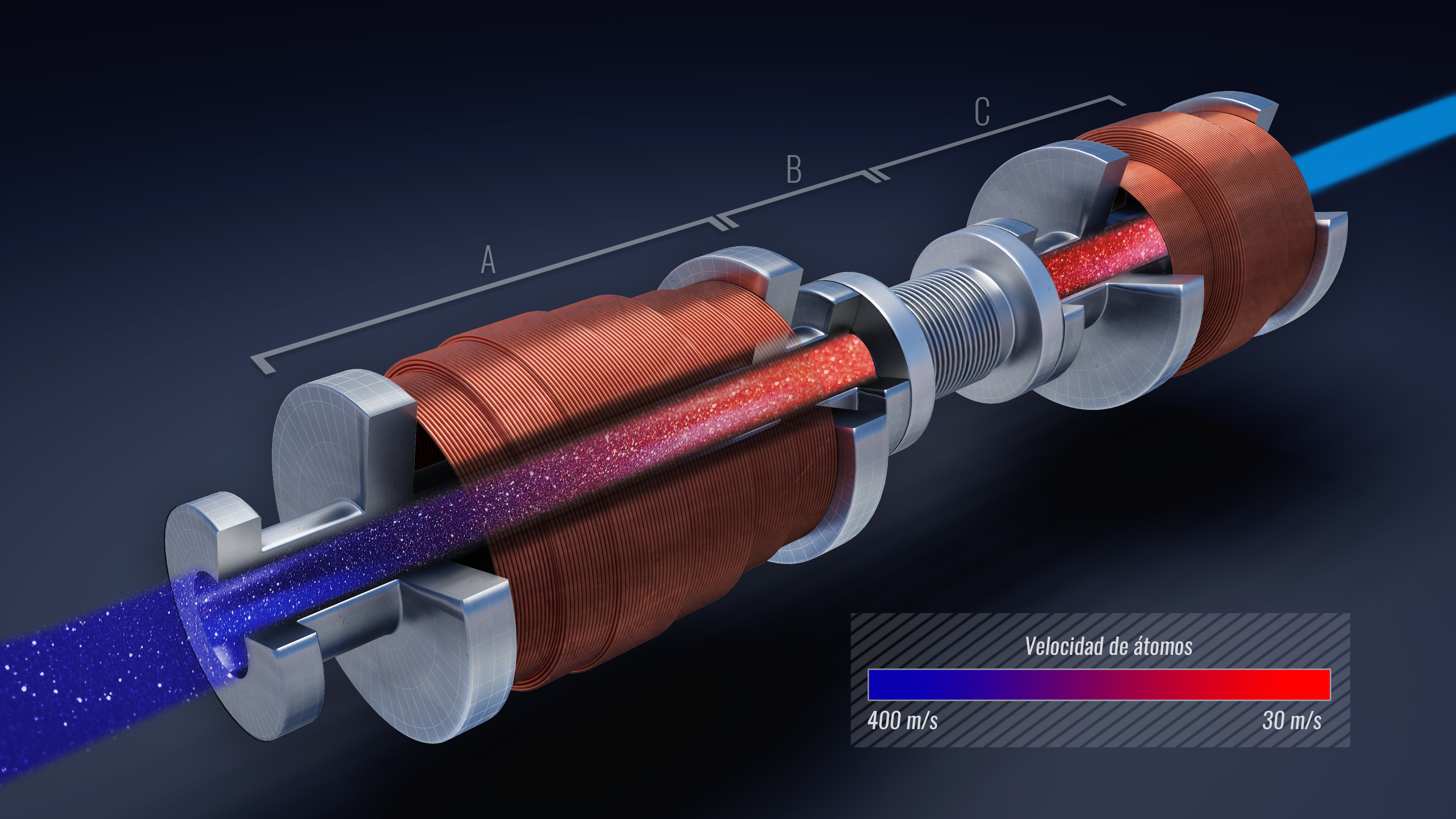}
\caption{Representaci\'on art\'istica del enfriamiento de \'atomos a trav\'es del desacelerador Zeeman. Un haz colimado de \'atomos es emitido con una rapidez de apr\'oximadamente $400\,\frac{\textrm m}{\textrm s}$, los \'atomos son frenados hasta alcanzar una rapidez de aproximadamente $30\,\frac{\textrm m}{\textrm s}$ por medio de la interacci\'on de la radiaci\'on electromagn\'etica de un l\'aser y un campo magn\'etico variable distribuido en tres secciones. La Secci\'on A) alcanza una intensidad de campo magn\'etico de un m\'aximo de -300\,G, la Secci\'on B) se mantiene en 0\,G y la Secci\'on C) alcanza un m\'aximo de +300\,G.}
\label{Sistema}
 \end{figure}

A partir de los resultados obtenidos con la aplicaci\'on del m\'etodo descrito en la Secci\'on~II se construy\'o el dise\~no propuesto para el enfriamiento de \'atomos de estroncio. La intensidad del campo magn\'etico en un punto dado es directamente proporcional al n\'umero de espiras o capas que rodean circularmente ese punto. Manteniendo una corriente constante, las secciones con mayor cantidad de capas generan un campo magn\'etico mayor y al disminuir el n\'umero de capas disminuye, consecuentemente, la intensidad del campo magn\'etico. El embobinado se realiz\'o con cable de cobre \# 13 ($1,845$\,mm de di\'ametro), sobre un tubo de cobre de 12,5\,cm de di\'ametro exterior y 1,5\,mm de espesor. Para contener las espiras se soldaron dos bridas para sellar metal de conexi\'on de bronce a cada extremo de 15\,cm de di\'ametro y 3\,cm de espesor. 

\begin{figure}
\centering
\includegraphics[height=3cm]{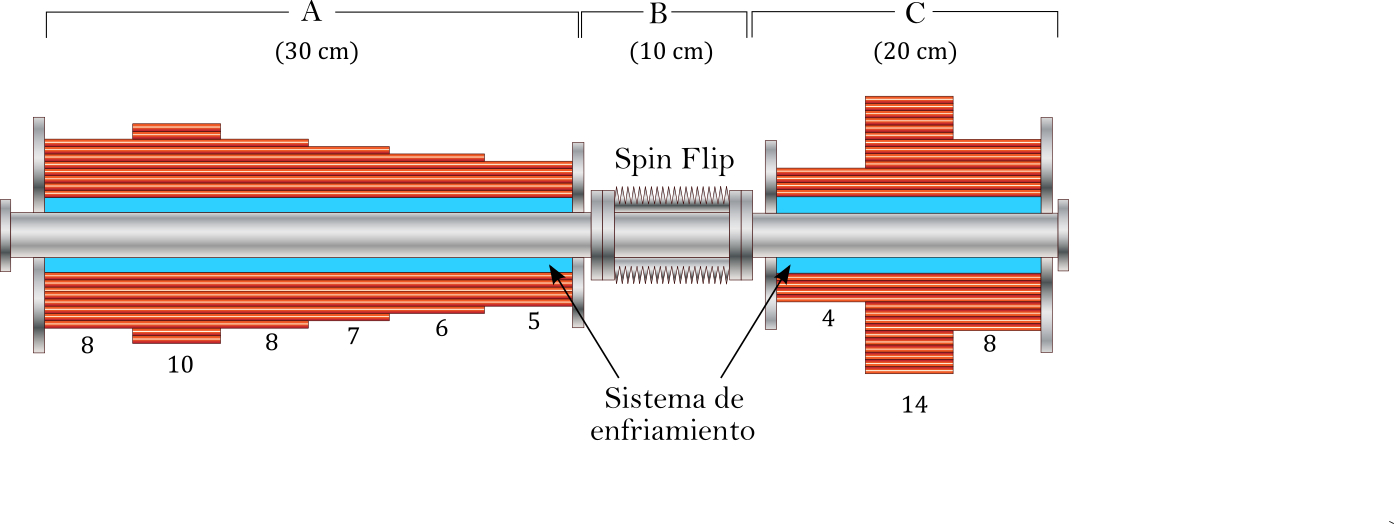}
\caption{Esquema del solenoide dise\~nado. Las secciones A, B y C corresponden a las indicadas en la Figura~\ref{Sistema}. Se indica la longitud de cada Secci\'on así como el n\'umero de capas en cada segmento del embobinado.}
\label{Esquema}
 \end{figure}

Para la obtenci\'on de la rapidez final de los \'atomos deseada, el desacelerador Zeeman se dise\~n\'o en tres secciones distintas  a lo largo del eje central: A) una Secci\'on de campo magn\'etico creciente, B) una Secci\'on de ``spin-flip'' o de campo magn\'etico cero y C) una Secci\'on de campo magn\'etico decreciente (ver Figura \ref{Sistema} y Figura~\ref{Esquema}) (un sistema similar se puede encontrar en Ref.~\cite{spinflip}). 
La Secci\'on A) tiene una longitud total de aproximadamente 25\,cm. Esta Secci\'on est\'a dise\~nada de forma tal que, al aplicar una corriente de 6\,A, la intensidad del campo magn\'etico inicia en 0\,G hasta alcanzar un valor m\'aximo negativo de -300\,G (1\,G=$10^{-4}$\,T) y luego decrezca hasta retornar a 0\,G. El embobinado de esta Secci\'on se realiz\'o en 6 segmentos de igual longitud (4\,cm) cuyo n\'umero de capas var\'ia en funci\'on del campo magn\'etico requerido. El primer segmento est\'a constituido por 8 capas, el segundo por 10 capas, el tercero tiene igualmente 8 capas y a partir de este en cada segmento el n\'umero de capas decrece en una unidad hasta llegar al \'ultimo segmento con 5 capas (ver sección A) de la Figura \ref{Esquema}). La Secci\'on B) est\'a constituida \'unicamente por una junta de expansi\'on de tipo fuelle de acero inoxiable de 10\,cm de largo esto para absorver cualquier vibraci\'on del sistema (ver sección B) de la Figura \ref{Esquema}). La Secci\'on C), con una longitud de 15\,cm, est\'a dise\~nada para alcanzar una intensidad de campo magn\'etico m\'axima de +300\,G y luego decrecer lentamente hasta 0\,G. En este caso el embobinado se reparti\'o en tres segmentos de igual longitud (5\,cm), el primero consta de 4 capas, el segundo 14 capas y el tercero 8 capas (ver sección C) de la Figura \ref{Esquema}). Para realizar el embobinado se fij\'o el tubo de cobre a un torno, asegurando as\'i que el cable fuera enrollado lo m\'as uniformemente posible. Durante todo el proceso de embobinado se control\'o que el cable no tuviera ning\'un contacto directo con el tubo de cobre para asegurar que no hubiera corto entre el cable y el tubo de cobre.

La potencia m\'axima generada por calentamiento de Joule debido a la corriente que circula por todo el sistema es de aproximadamente 160\,W. Para controlar el calentamiento generado por esta potencia se incorpor\'o un sistema de enfriamiento por medio de circulaci\'on de agua. El anterior est\'a constituido por un tubo de cobre de un di\'ametro exterior menor de 10\,cm y 1,5\,mm de espesor, soldado a las bridas de bronce de conexi\'on del embobinado. A este tubo se acoplan dos conectores en los extremos, los cuales son conectados a un flujo constante de agua. Esto permite mantener una c\'amara de enfriamiento, que separa el embobinado de la secci\'on interna por la cual viajan los \'atomos, asegurando una temperatura inferior a $30\,^{\circ}$C.

La rapidez de los \'atomos es reducida aleatoriamente por medio de la interacci\'on de la radiaci\'on electromagn\'etica de un l\'aser y el campo magn\'etico de las capas, en el interior de un tubo de acero inoxidable de 8\,cm de di\'ametro y 15\,mm de espesor que pasa por el centro del desacelerador Zeeman. Esta secci\'on se acopla al sistema de vac\'io por medio de dos bridas de conexi\'on de acero inoxidable de 10\,cm de di\'ametro y 2\,mm de espesor. 

Para realizar las mediciones de campo magn\'etico se utiliz\'o un sensor de intensidad de campo magn\'etico Vernier Magnetic Field\,$^{\textregistered}$ modelo MG-BTA cuya m\'inima divisi\'on de escala es de 0,001\,G y cuyo rango de medici\'on es de 0\,G hasta $\pm80$\,G. Las mediciones de temperatura se realizaron con un termopar de tipo T (cobre-constantan) cuyo rango de medici\'on es de $-200\,^{\circ}$C a $260\,^{\circ}$C y sensibilidad es de $43\ \frac{\mu \textrm{V}}{^{\circ}\textrm{C}}$. Se realizaron mediciones de presi\'on con un sensor modelo MDC Nude Ionization Gauge Tube$^{\textregistered}$ cuyo rango m\'aximo de medici\'on es de $1 \times 10^{-3}$\,Torr hasta  $2 \times 10^{-10}$\,Torr con una sensibilidad del 20\,\%.

\section{Comparaci\'on entre simulaci\'on y datos experimentales}

\begin{figure}
\centering
\includegraphics[height=5cm]{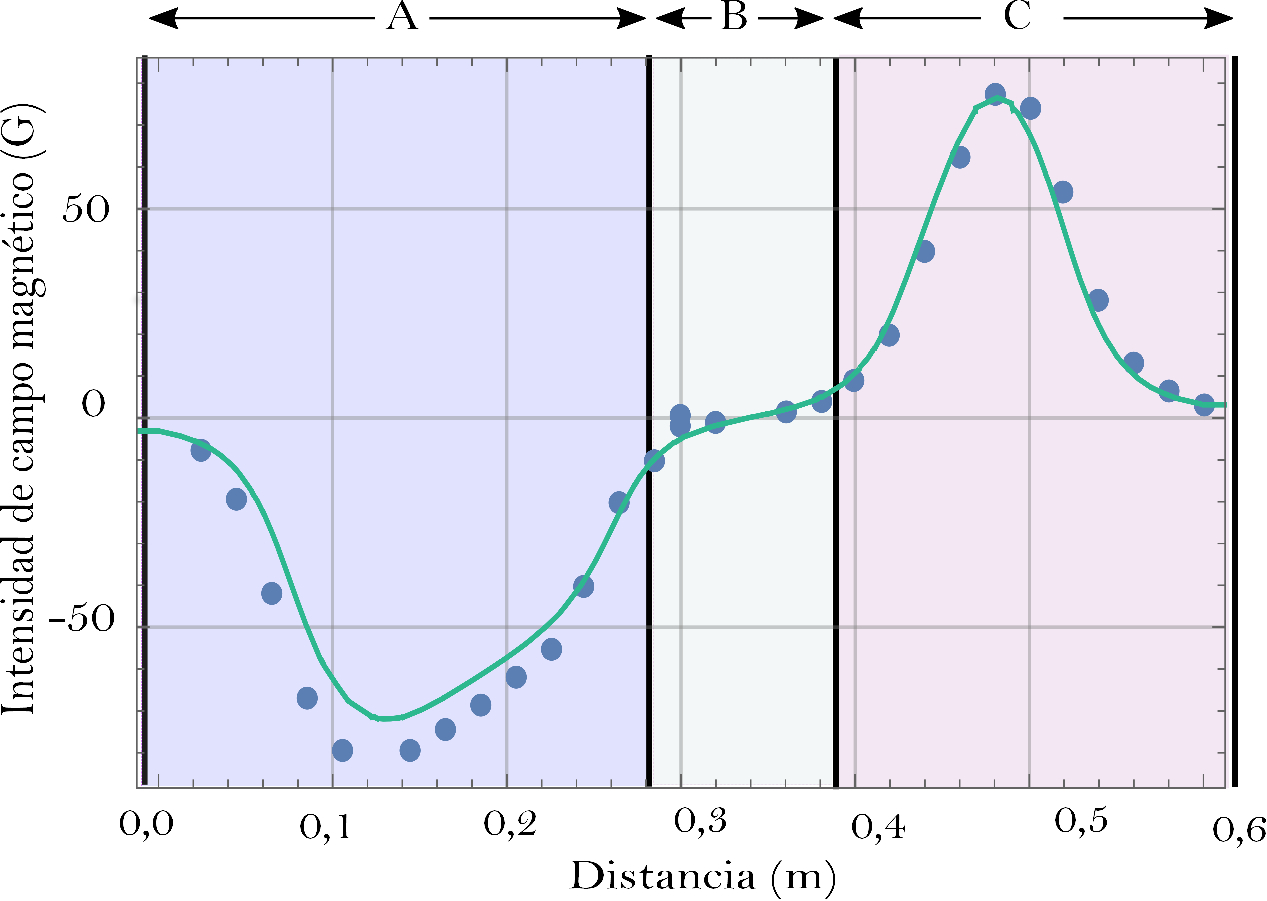}
\caption{Comparaci\'on de la intensidad de campo magn\'etico medida en el desacelerador Zeeman en funci\'on de la distancia con los valores determinados por el modelo descrito en la Secci\'on II utilizando una corriente de 1,5\,A. Los puntos experimentales corresponden a los valores de la intensidad de campo magn\'etico medidos } y la l\'inea s\'olida corresponde a la variaci\'on de la intensidad del campo magn\'etico en funci\'on de la distancia generada por el modelo. Las l\'ineas negras delimitan las secciones del embobinado. Las partes sombreadas corresponden a las tres secciones descritas en la Figura\,\ref{Sistema} y en la Figura\,\ref{Esquema}.
\label{campo1A}
 \end{figure}

Para corroborar los modelos descritos anteriormente en la Secci\'on II se realizaron mediciones de la intensidad del campo magn\'etico en funci\'on de la distancia a trav\'es del desacelerador Zeeman. Las mediciones de intensidad del campo magn\'etico se llevaron a cabo utilizando una corriente de 1,5\,A con la cual se alcanza, seg\'un el modelo, un campo m\'aximo de $\pm80$\,G (ver Figura \ref{campo1A}), lo anterior debido a que este valor concuerda con el rango de medici\'on del sensor utilizado. Se realizaron tres repeticiones de la mediciones y se obtuvieron coeficientes de variaci\'on inferiores al 2\,\% con lo cual se comprueba la estabilidad del campo. La Figura \ref{campo1A} muestra la buena conformidad entre el modelo descrito en la Secci\'on II y los resultados experimentales. Las diferencias encontradas no superan el $7.5\%$, este valor, a pesar de ser elevado, es esperado debido a las dificultades presentes al realizar el embobinado de forma uniforme. Los datos experimentales muestran en la Figura \ref{campo1A} que, efectivamente, en la Secci\'on A) la intensidad de campo magn\'etico inicia en 0\,G y disminuye hasta alcanzar un m\'aximo negativo de $-80$\,G y luego crece hasta alcanzar un valor de 0\,G. En la Secci\'on B) el campo se mantiene cercano a 0\,G. Finalmente, la Secci\'on C)  mantiene igualmente la tendencia deseada, un incremento hasta un m\'aximo de $+80$\,G seguido por una disminuci\'on del campo magn\'etico hasta alcanzar un valor cercano a $0$\,G.

Los resultados obtenidos nos permiten, adem\'as, corroborar las ventajas del m\'etodo desarrollado en esta investigaci\'on. Se observa, en la Figura \ref{campo1A}, como la simulaci\'on por medio de elementos finitos permite predecir con mucha precisi\'on el comportamiento del campo magn\'etico generado por el dise\~no del desacelerador Zeeman. Lo anterior permite hacer cualquier ajuste que sea necesario en el dise\~no antes de llevar a cabo su construcci\'on. Por otro lado, otras investigaciones utilizan \'unicamente un modelo matem\'atico para determinar la configuraci\'on del embobinado, como es reportado en Ref.~\cite{zeemanslower, tesis}. A pesar de que reportan resultados satisfactorios presentan la necesidad de implementar modificaciones en el dise\~no construido, tales como, variaciones en el embobinado, implementaci\'on de m\'as bobinas o de un escudo magn\'etico. El met\'odo que proponemos tiene la ventaja de que permite tener una gran confianza con respecto al perfil de intensidades del campo magn\'etico del dise\~no a construir.

Para comprobar que la intensidad del campo magn\'etico en el exterior del desacelerador Zeeman tiende a 0\,G conforme se aleja de los extremos (inicio de Secci\'on A) y final de Secci\'on C)),  se realizaron mediciones de la intensidad del campo magn\'etico en funci\'on de la distancia al alejarse de ambos extremos del desacelerador Zeeman. En ambos casos se comprob\'o que la intensidad del campo magn\'etico cerca de los extremos alcanza un valor m\'aximo de aproximadamente 5\,G y cae r\'apidamente a valores cercanos a 0\,G al alcanzar una distancia de 5\,cm de ambos extremos. Estos resultados son conformes con la simulaci\'on realizada, lo cual, permite confirmar una vez m\'as las ventajas del m\'etodo propuesto. 

Para el caso de reducir la rapidez de los \'atomos de estroncio alrededor de un orden de magnitud es necesario suministrarle al desacelerador Zeeman una corriente de 6\,A para obtener, seg\'un el modelo desarrollado, intensidades de campo magn\'etico de $\pm300$\,G. Debido a las limitaciones del sensor de intensidad de campo magn\'etico utilizado es necesario comprobar qu\'e tipo de relaci\'on existe entre el incremento de la corriente y la intensidad del campo magn\'etico obtenido. Por lo tanto, se realizaron mediciones de la variaci\'on del campo en un punto espec\'ifico en funci\'on de la corriente y se observa, efectivamente, un comportamiento lineal, la intensidad del campo magn\'etico incrementa proporcionalmente al aumento de la corriente. Se puede afirmar que, al utilizar una corriente de 6\,A se obtiene una intensidad de campo magn\'etico 4 veces mayor que el medido con una corriente de 1,5\,A. Estos valores concuerdan con el perfil de campo magn\'etico necesario para realizar el enfriamiento de los \'atomos de estroncio. 

\begin{figure}
\centering
\includegraphics[height=5cm]{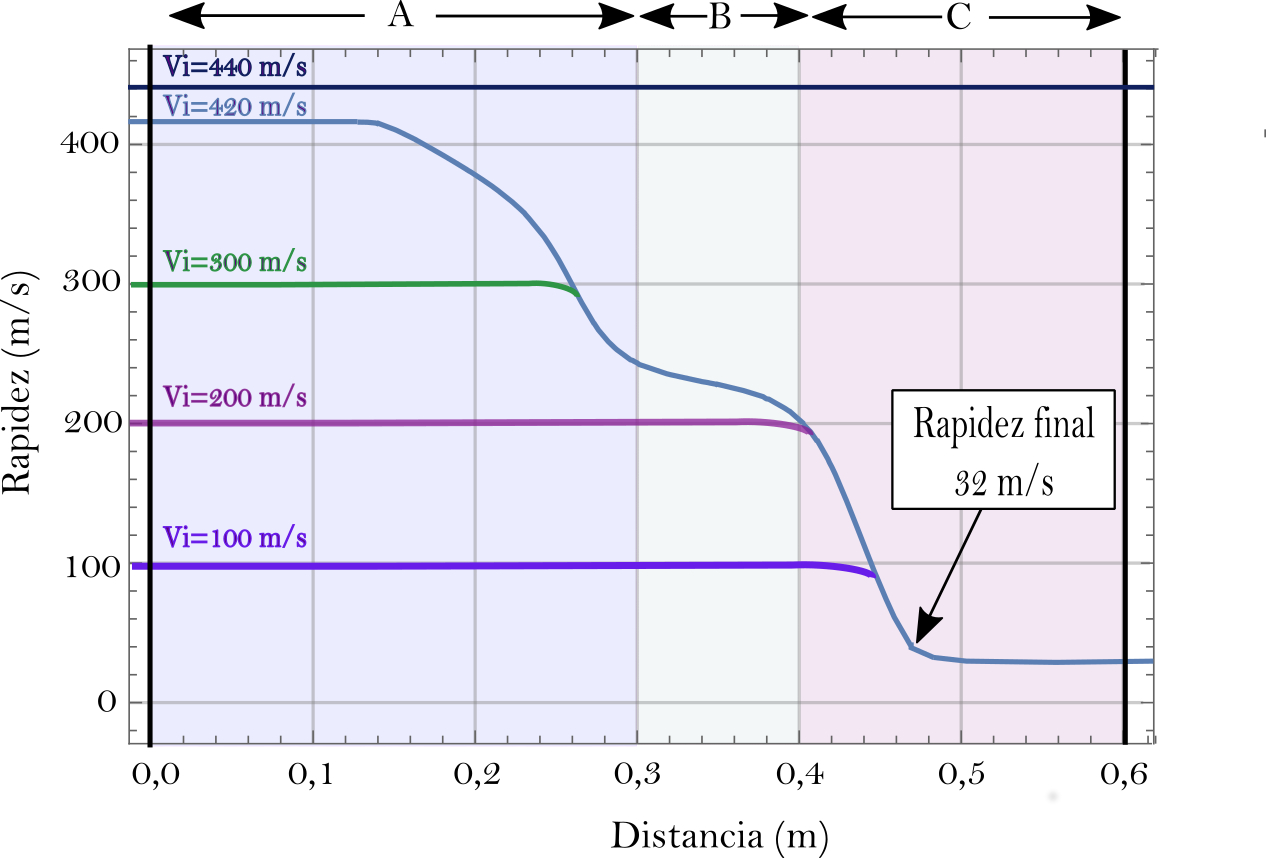}
\caption{Rapideces de los \'atomos en funci\'on de la distancia en el desacelerador Zeeman, utilizando una corriente de 6\,A, para diferentes rapideces iniciales. Las l\'ineas negras delimitan la Secci\'on del embobinado. Las partes sombreadas corresponden a las tres secciones descritas en la Figura \ref{Sistema} y en la\ref{campo1A}.}
\label{velocidad}
 \end{figure}

Para poder predecir c\'ual va a ser la variaci\'on de la rapidez de los \'atomos se utilizan las mediciones de la variaci\'on de la intensidad del campo magn\'etico generada por el desacelerador Zeeman utilizando una corriente de 6\,A en funci\'on de la distancia \cite{zeemanslower}. A partir de la Ecuaci\'on \ref{eq:1} se deduce la variaci\'on de la rapidez de los \'atomos en funci\'on de la intensidad de campo magn\'etico.

La Figura \ref{velocidad} muestra la variaci\'on de la rapidez de los \'atomos en funci\'on de la posici\'on en el desacelerador Zeeman para diferentes rapideces iniciales, es decir a la velocidad \textsl{$v_{0}$} en la Ecuaci\'on~\ref{eq:2}, utilizando una corriente de 6\,A. El sistema est\'a dise\~nado para frenar \'atomos con una rapidez inicial m\'axima de $420\,\frac{\textrm{m}}{\textrm{s}}$. \'Atomos con rapideces iniciales inferiores empiezan a ser frenados m\'as adelante en la Secci\'on de enfriamiento (ver Figura \ref{velocidad}), esto ocurre a partir del momento en que entran en resonancia con la frecuencia del l\'aser. Los \'atomos son frenados en la Secci\'on C) del desacelerador Zeeman a la cual llegan con una rapidez final de aproximadamente $32\,\frac{\textrm{m}}{\textrm{s}}$ .

Para comprobar el funcionamiento del sistema de enfriamiento, se realizaron mediciones de la temperatura en distintas secciones del desacelerador Zeeman. Se observaron las variaciones de la temperatura en funci\'on de la corriente aplicada con y sin el flujo de agua. Se observ\'o que, cuando el sistema de enfriamiento no est\'a activado, se alcanza una temperatura m\'axima de 98\,$^{\circ}$C, al utilizar una corriente de 10\,A despu\'es de 15 minutos y cuando se activa el sistema la temperatura no incrementa de 62\,$^{\circ}$C despu\'es de 15 minutos. 

\section{Conclusi\'on}

En esta investigaci\'on se present\'o el desarrollo de un nuevo m\'etodo para optimizar el dise\~no de un desacelerador Zeeman. Particularmente este m\'etodo fue implementado en la realizaci\'on de un sistema solenoidal de capas variables para un desacelerador Zeeman de tipo ``Spin-Flip''. El m\'etodo propuesto est\'a constituido por dos modelos mat\'ematicos. El primer modelo permite obtener una configuraci\'on del solenoide y el segundo realiza una simulaci\'on por medio de elementos finitos.  Se demostr\'o que al utilizar el segundo modelo es posible predecir con una muy alta precisi\'on el comportamiento del perfil de intensidad del campo magn\'etico generado por el desacelerador Zeeman. Para lograr el objetivo mencionado anteriormente, se busc\'o, por un lado generar  un perfil de intensidades lo m\'as cercano posible al perfil te\'orico y que por otro lado asegurara que la intensidad del campo magn\'etico tendiera a cero en el exterior del desacelerador Zeeman. Con la informaci\'on anterior, se construy\'o un desacelerador Zeeman con lo cual se comprob\'o que el perfil de intensidades de campo magn\'etico generado concuerda con las predicciones de la simulaci\'on y se ajusta al modelo te\'orico. Se implement\'o, adem\'as, un sistema de enfriamiento el cual, por medio de circulaci\'on de agua, permiti\'o contrarrestar en un 40\% el calentamiento generado por el efecto Joule. 

A partir de las mediciones de las variaciones de la intensidad del campo magn\'etico se realiz\'o un modelo que permite predecir las variaciones de la rapidez de los \'atomos a lo largo del desacelerador Zeeman. En un trabajo que est\'a por publicarse \cite{publicarse}, se comprobar\'an las predicciones del modelo descrito en esta investigaci\'on por medio de mediciones de la rapidez final de los \'atomos, una vez que han atravesado el desacelerador Zeeman, en funci\'on de la rapidez inicial. 

\section{Agradecimientos}

Nosotros estamos muy agradecidos por la dedicaci\'on y colaboraci\'on de V\'ictor Araya Rodr\'iguez en la construcci\'on del desacelerador Zeeman y a Felipe Molina Guti\'errez por su valiosa colaboraci\'on en la elaboraci\'on art\'istica de las figuras presentadas en el art\'iculo. Nosotros tambi\'en queremos agradecer a la Vicerrector\'ia de Investigaci\'on de la Universidad de Costa Rica por el apoyo brindado. L.S. agradece el apoyo dado por el DAAD (Deutscher Akademischer Austauschdienst) a través del programa RISE (Research Internship in Science and Engineering).

\end{document}